\documentclass[useAMS,usenatbib]{mn2e}
\usepackage{amsmath}
\usepackage{amsbsy}
\usepackage{times}
\usepackage{graphicx}
\usepackage{aas_symbols}
\usepackage{epstopdf}
\usepackage{cases}
\usepackage{color}

\newcommand{\beq}{\begin{equation}}
\newcommand{\eeq}{\end{equation}}
\newcommand{\bea}{\begin{eqnarray}}
\newcommand{\eea}{\end{eqnarray}}
\newcommand{\gae}{\lower 2pt \hbox{$\, \buildrel {\scriptstyle >}\over {\scriptstyle
\sim}\,$}} 
\newcommand{\lae}{\lower 2pt \hbox{$\, \buildrel {\scriptstyle <}\over {\scriptstyle
\sim}\,$}}

\begin{document}

\title[Off-axis afterglows and the accompanying SN]{GRB off-axis afterglows and the emission from the accompanying supernovae}

\author[Kathirgamaraju, Barniol Duran \& Giannios]{Adithan Kathirgamaraju$^{1}$\thanks{Email: akathirg@purdue.edu (AK), rbarniol@purdue.edu (RBD), dgiannio@purdue.edu (DG)}, Rodolfo Barniol Duran$^{1}$\footnotemark[1], Dimitrios Giannios$^{1}$\footnotemark[1] \\
$^{1}$Department of Physics and Astronomy, Purdue University, 525 Northwestern Avenue, West Lafayette, 47907 IN, USA
}

\date{Accepted; Received; in original form 2016 March 18}

\pubyear{2016}

\maketitle

\begin{abstract}
Gamma-ray burst (GRB) afterglows are likely produced in the shock that is driven as the GRB jet interacts with the external medium. Long-duration GRBs are also associated with powerful supernovae (SNe). We consider the optical and radio afterglows of long GRBs for both blasts viewed along the jet axis (``on-axis'' afterglows) and misaligned observes (``off-axis'' afterglows). Comparing the optical emission from the afterglow with that of the accompanying SN, using SN 1998bw as an archetype, we find that only a few percent of afterglows viewed off-axis are brighter than the SN. For observable optical off-axis afterglows, the viewing angle is at most twice the half-opening angle of the GRB jet.  Radio off-axis afterglows should be detected with upcoming radio surveys within a few hundred Mpc. We propose that these surveys will act as ``radio triggers," and that dedicated radio facilities should follow-up these sources. Follow-ups can unveil the presence of the radio SN remnant, if present.  In addition, they can probe the presence of a mildly relativistic component, either associated with the GRB jet or the SN ejecta, expected in these sources. 
\end{abstract}


\section{Introduction}

Gamma-ray burst (GRB) afterglows are likely to be produced in the external forward shock (e.g., \citealp{sarietal98, wijersandgalama99, panaitescuandkumar00}).  In this framework, the GRB jet interacts with the external medium and drives a relativistic shock, accelerating electrons that radiate via synchrotron emission.  The decelerating blast wave, initially highly collimated, transitions from a relativistic stage to a non-relativistic spherical stage at late times. Hydrodynamical simulations are, nowadays, able to capture this long-term evolution of the blast wave and calculate multi-wavelength synchrotron light curves and spectra (e.g., \citealp{vaneertenandmacfadyen12}), for observers located along the jet axis (``on-axis" observers) and at a large angle (``off-axis" observers; e.g., \citealp{rhoads1997, granotetal02, totani2002, nakar2002, zou2007, rossi2008}). Off-axis afterglows can potentially be observed without the detection of the prompt gamma-ray emission. For this reason, they have been referred to as ``orphan afterglows". The detection of orphan afterglows remains elusive to this date (e.g.,  \citealp{cenko2013, corsi2015}). Current and upcoming surveys in the optical (e.g., Pan-STARRS1, ZTF, LSST) and radio (e.g., LOFAR, VAST, VLASS, SKA1) have the detection of orphan afterglows among their main objectives.  

Long GRBs are also accompanied by supernovae (SNe) of the rare broad-line Ic type.  The sample of GRB-associated SNe is quite homogeneous, and the optical SN emission from SN 1998bw serves as an excellent archetype (e.g., \citealp{modjaz}).  Modelling of the SN optical emission reveals typically very energetic ejecta with kinetic energy of several $\times 10^{52}$ erg, and fairly fast velocity of $\sim 0.1$c (see, e.g, \citealp{woosleyandbloom06, hjorthandbloom12, melandrietal14}, and references therein). The SN ``remnant" also drives an external shock, accelerating electrons that radiate via synchrotron emission (e.g., \citealp{chevalier82a, chevalier82b, chevalier98}).  Recently, the emission from this SN remnant (SNR) has been shown to produce a strong radio signal that could potentially be observed $\sim 10$ yrs after the GRB explosion \citep{barniolduranandgiannios15}.

Typical GRBs occur at cosmological distances. Current and near-future facilities will be capable of detecting orphan afterglows from much closer distances: just $\sim$ a few hundred Mpc. Therefore, future orphan afterglow observations should increase the number of GRBs detected nearby.  This carries the promise of following these afterglows for decades in the radio, and studying the very late stages of the shock, including the potential detection of the onset of the SNR emission. 

The main objective of this paper is to calculate the emission that follows the GRB, including the afterglow and the SNR, for an observer located at any angle with respect to the jet axis. We assess the various strategies of detecting an orphan afterglow at different wavelengths, especially optical and radio.  In particular, in the optical band, we compare the expected afterglow emission with the optical emission from the SN itself.  We stress, in line with previous work, that radio frequencies constitute the best observing strategy to detect the emission from an orphan afterglow (e.g., \citealp{paczynski01, levinsonetal02, bergeretal03, soderbergetal04, galyam2006, soderbergetal06, bietenholz2014, ghirlanda2014, metzger2015}).  

It is likely that the two components that we have mentioned above, the GRB jet and the SN ejecta, are not simply expelled from the central object with a single velocity and a characteristic energy.  Instead, the quasi-spherical SN ejecta is thought to be composed of a range of energies that follow a power-law in velocity, with faster parts of the ejecta carrying smaller energies, as expected in hydrodynamical explosions (e.g., \citealp{matzner1999}).  At the same time, a distribution of energies could also be present in the GRB jet (e.g.,
\citealp{lazzatietal12}), or the GRB jet could be surrounded by a slower ``sheath" (or ``cocoon") of a larger opening angle (e.g., \citealp{ramirez2002}; \citealp{weiqun2004}). These possibilities motivate us to consider the presence of a mildly relativistic component in the ejecta.  In this paper we calculate the radiative signatures of such a component.

The paper is organized as follows. In Section \ref{Section2} we describe the emission from the different components that follow the GRB prompt emission. In Section \ref{Section_results} we present optical and radio light curves of these components. In Section \ref{Discussion} we comment on their potential detection, and briefly discuss observing strategies and rates.  We finish with our conclusions in Section \ref{Conclusions}.

\section{Modelling the emission that follows the GRB} \label{Section2}

The long-lasting emission that follows the prompt GRB emission has different components: (i) the external forward shock emission that is initially strongly beamed along the direction of propagation of the jet but that gradually turns spherical as the blast slows down; (ii) the quasi-spherical SN optical emission powered by the radioactive decay and (iii) the quasi-isotropic SNR emission, which is produced by synchrotron emission from electrons accelerated at the SN shock. For completeness, we explore the possible contribution of synchrotron emission from an external forward shock driven by mildly relativistic ejecta.  We discuss all of these in the following subsections.

\subsection{GRB jet afterglow model} 

\subsubsection{Afterglow library}

We calculate the GRB afterglow light curves using the ``Afterglow library" described in \cite{vaneertenandmacfadyen12} (see, also, \citealp{zhangandmacfadyen09, vaneertenandwijers09, vaneertenetal10a, vaneertenetal10b,vaneertenandmacfadyen11}).\footnote{ The Afterglow library (here we use the ``BOXFIT" code) is publicly available at http://cosmo.nyu.edu/afterglowlibrary.} The library calculates the synchrotron light curves and spectra (at a given frequency and for a given observer angle with respect to the jet axis) using linear radiative transfer, which includes synchrotron self-absorption. The library uses snapshots of hydrodynamical simulations of GRB jets to generate these light curves. In this paper, we modify the library as described in \cite{sironiandgiannios13}.  This modification allows us to consider the ``deep Newtonian (DN)" regime, which is relevant for the late-time light curves calculation, where most of the shock-heated electrons are non-relativistic, but mildly relativistic particles with Lorentz factor $\sim 2$ contribute to the bulk of the total electron energy (see, also, \citealp{granot2006}).

\subsubsection{Baseline: Optical and radio light curve on-axis modelling} \label{Baseline_optical}

We use the sample of optical on-axis GRB afterglow observations found in \cite{kannetal10} as a baseline for our study. This comprehensive sample has been extinction-corrected and scaled to a common redshift $z=1$ and common R-band ($\sim 2$ eV) magnitude. We arrange these observed afterglows in descending order of their brightness at 1 day and then divide them into 10 groups. The first group contains the 10\% brightest afterglows ($9^{th}$ decile), the second group contains the subsequent 10\% brightest afterglows ($8^{th}$ decile) and so on. For each decile, we produce an average optical afterglow light curve so that we can use it to represent that decile. We also calculate the decile's average isotropic gamma-ray energy, $E_{\rm \gamma,iso}$, which is the energy released during the prompt GRB phase. 

To model the representative afterglow light curve of each decile, we need several parameters: $E_{\rm iso}$, the isotropic equivalent kinetic energy of the jet; $n$, the number density of the external medium (assumed constant); the microphysical parameters $\epsilon_e$ and $\epsilon_B$, the fraction of energy in the electrons and magnetic field in the shocked fluid, respectively; and $p$, the power-law index of the electron energy distribution. Although modelling of afterglow data of different GRBs indicates that these parameters are not universal, we first assume a particular set of values (based on recent studies), and later discuss how varying some of these parameters would affect our results (see Section \ref{Discussion} and the Appendix).

For each decile, we assume a GRB gamma-ray efficiency of $\sim$20\%, so that $E_{\rm iso} \approx 5 E_{\rm \gamma,iso}$ (e.g., \citealp{beniamini2015}), $\epsilon_{e} \sim 0.1$ (\citealp{santana2014}), $p = 2.4$ (\citealp{curran2010}). We assume $n=1$ cm$^{-3}$, and then find a suitable value of $\epsilon_{B}$ to match the average optical brightness at 1 day for each decile using the Afterglow library, which turns out to be $\sim 10^{-5}-10^{-4}$ consistent with, e.g., \cite{barniolduran2014}, \cite{santana2014} and \cite{beniamini2015}. A large number of afterglow studies seem to point out that $\epsilon_e$ is quite constrained to be $\sim 0.1$ (e.g., \citealp{santana2014}, see, also, particle-in-cell simulations of \citealp{sironi2011}). It also seems that the prompt gamma-ray efficiency should not be too high (e.g., \citealp{beniamini2015}). Hence, we investigate below how changing the remaining parameters $n$ and correspondingly $\epsilon_B$, which seem to be the least constrained, affects our conclusions. To illustrate this, we can analytically estimate the optical flux at 1 day, when the optical band is likely to be above the minimum frequency, but below the cooling one. It is given by (e.g., \citealp{granotandsari02})
\begin{equation}
F_{\nu} \approx (3 {\rm mJy}) \epsilon_{\rm e,-1}^{1.4} \epsilon_{\rm B,-4}^{0.9} E_{\rm iso,53}^{1.4} n_0^{1/2} t_d^{-1}d_{\rm 27}^{-2} \nu_{\rm 14}^{-0.7},
\end{equation}
where we have used the parameters mentioned above, $t$ is the observed time since the explosion (in days), we have normalized the luminosity distance $d$ to 300 Mpc, and we have used the common notation $Q_x = Q/10^x$ in c.g.s units. Since we consider nearby sources, we take the redshift to be $1+z\approx 1$ in our equations. For example, for the 5$^{th}$ decile, $E_{\rm iso,53} \sim 2$, the flux at 1 d (at 300 Mpc) at 2 eV is $\sim 5$ mJy, and therefore, $n=1$ cm$^{-3}$ requires $\epsilon_B \sim 4\times10^{-4}$. This value of $\epsilon_B$ agrees with the one found using the Afterglow library within a factor of $\sim 2$.

The final parameter needed is the half-opening angle of the jet, $\theta_j$, which affects the time of the ``jet break" (e.g., \citealp{rhoads1999}; \citealp{sari1999}). Individual fitting of each of the light curves is needed to do this, which is outside of the scope of this paper. Therefore, we use a simple approach and consider $\theta_j=0.1$ ($\sim 6^{\circ}$) and $0.2$ ($\sim 11^{\circ}$), which spans the approximate range of typical opening angles inferred from observations (e.g., \citealp{ghirlanda2005}; \citealp{goldstein2016} and references therein). We now introduce some terminology that will be used throughout this paper. The afterglows in the $9^{th}$ decile will be called the ``brightest afterglows." We also group the GRBs in the $4^{th}$, $5^{th}$ and $6^{th}$ deciles together and call these the ``average afterglows," since they yield the average optical flux at 1 d of our sample.

Using the parameters described above, we predict the off-axis afterglow in the optical (R-band). We also use these parameters to predict the on-axis and off-axis radio afterglows (at 4.9 GHz, which is a typical observing radio band). Given that radio on-axis afterglows observations are available, we use the average observed radio afterglow light curve in \cite{chandra2012} to compare with our predicted on-axis radio afterglows.

\subsection{SN optical emission} \label{SN_section}

The SNe accompanying long duration GRBs are very similar in nature. At $\sim 10$ d after the GRB, the GRB-SNe sample spans only a factor of $\lae 4$ in bolometric luminosity (\citealp{melandrietal14}). Also, in a systematic study recently done by \cite{modjaz}, it was found that SN 1998bw represents a typical SN that accompanies a GRB. Hence, we will use the derived physical parameters of 1998bw in this paper: an SN ejecta kinetic energy of $E_{\rm SN} \sim 5 \times 10^{52}$ erg and a velocity of 24,000 km/s, which is $\beta_{\rm SN} \approx 0.08$ in units of the speed of light (e.g., \citealp{iwamoto1998}). The optical data for SN1998bw were obtained from \cite{SN1998bw}.

\subsection{SN remnant (SNR) radio emission} \label{SNR_section}

At very late times, radio emission from the SNR may outshine the GRB afterglow (\citealp{barniolduranandgiannios15}). Therefore, it proves useful to include this component in our late time calculations. We follow the same procedure as in \cite{barniolduranandgiannios15} to obtain the SNR light curve for a 1998bw-like SN. To summarize their work, the SNR radio flux is 
\begin{subnumcases}{\label{SNRflux} F_{\nu} = F_p} 
t^{3} & $t <  t_{\rm dec,SN}$\label{before}\\
t^{\frac{-3(1+p)}{10}} & $t > t_{\rm dec,SN}$, \label{after}
\end{subnumcases}
where the deceleration time of the SN ejecta is 
\beq
\label{tdec}
t_{\rm dec,SN} \approx 29 \beta_{\rm SN,-1}^{-5/3} (E_{\rm SN,52.5}/n_0)^{1/3} \, \rm yr,
\eeq
and the peak flux (in $\mu$Jy) at observed frequency $\nu$ at this time is given by
\beq
F_p \approx 440 \, \bar{\epsilon}_{e,-1} \epsilon_{\text{B-SN},-2}^{\frac{1+p}{4}} \beta_{\rm SN,-1}^{\frac{1+p}{2}} 
E_{\rm SN,52.5} \, n_0^{\frac{1+p}{4}} \nu_{\rm GHz}^{\frac{1-p}{2}} d_{\rm 27}^{-2},
\label{F_DN_SN}
\eeq
where $\bar{\epsilon}_e \equiv 4 \epsilon_e (p-2)/(p-1)$. It is important to note that these results are valid for max($\nu_a,\nu_m$) $< \nu < \nu_c$, where $\nu_a$, $\nu_m$ and $\nu_c$ are the synchrotron self-absorption, minimum injection and cooling frequencies, respectively. We assume that the SNR emission is quasi-isotropic; therefore, it is observable for any viewing angle. 

Since SNR emission from a GRB-accompanying SN has not been observed, the SNR physical parameters remain uncertain. In the following, we will use the same values for density, $\epsilon_e$ and $p$ used for our GRB afterglows calculations, and we will use the kinetic energy and velocity of the SN ejecta inferred from 1998bw (see Section \ref{SN_section}). We will use a value of $\epsilon_{\text{B-SN}} \sim 0.01$, inferred for ``normal" Ibc young radio SNe (e.g. \citealp{chevalier2006}). Ideally, one could use very late time radio afterglow observations ($\sim 10$ yr) to constrain $\epsilon_B$ and other parameters, and use them to calculate the radio SNR emission. This can only be done for one GRB, so here we simply adopt a fixed value of $\epsilon_B$ (see \citealp{barniolduranandgiannios15}). A different choice of parameters would yield different fluxes according to equation (\ref{F_DN_SN}).

\subsection{Other possible mildly relativistic components} \label{Mildly_relativistic_section}

So far we have considered the light curves from two components with distinct energies and distinct velocities: the ultra-relativistic jet and the non-relativistic SN ejecta. Motivated by previous work, we consider the presence of a mildly relativistic component (e.g., \citealp{lazzatietal12}) and predict its light curve. This extra component could be related to the GRB jet or the SN ejecta.  For simplicity, we will assume that it is associated with the quasi-spherical SN ejecta. We will model the SN ejecta with a continuous kinetic energy distribution as $E\propto(\beta\gamma)^{-\alpha}$ with $0.1\leq\beta\gamma$, $\alpha$ varying from 1 to 5 (observations and theory seem to constrain $\alpha\sim5$ for typical SNe), and we normalize the total kinetic energy to that of the SN (e.g., \citealp{matzner1999}, \citealp{tan2001}). Faster parts of the ejecta, which contain less energy, decelerate at earlier times; slower but more energetic parts catch up with them later on, re-energizing the blast wave. In this scenario, the Lorentz factor and energy of the blast wave are $\Gamma \propto t^{-3/(8+\alpha)}$ and $E \propto t^{3\alpha/(8+\alpha)}$ (e.g., \citealp{sari2000}), while in the non-relativistic phase the velocity and energy are $\beta \propto t^{-3/(5+\alpha)}$ and $E \propto t^{3\alpha/(5+\alpha)}$. The synchrotron emission from this blast wave yields light curves of the form $F_{\nu}\propto t^{-s}\nu^{-(p-1)/2}$, with the temporal decay index presented in \cite{barniolduran15}. During the non-relativistic phase, the blast wave will transition to the DN phase.  The DN phase sets in when the minimum Lorentz factor of the shocked electrons drops to $\gamma_{min}\lesssim 2$ (\citealp{sironiandgiannios13}). At this stage, the mildly relativistic electrons contribute to the majority of the flux emitted, causing a slight change in the slope of the light curves.  The temporal decay indices in order of appearance are 
\begin{subnumcases}{\label{s} s=}
\frac{6(p-1)-3\alpha}{8+\alpha}&\text{relativistic phase,}\label{relativistic}\\
\frac{15p-21-6\alpha}{10+2\alpha}&\text{non-relativistic phase,}\label{nonrelativistic}\\
\frac{3(1+p)-6\alpha}{10+2\alpha}&\text{deep Newtonian phase.}\label{deepnewtonian}    
\end{subnumcases}
The transition from the relativistic to the non-relativistic phase occurs at $\beta \gamma = 1$, and the transition from the non-relativistic to DN phase occurs at a velocity $\beta_{\rm DN} = 0.2 {{\bar\epsilon_{e,-1}}}^{-1/2}$ \citep{sironiandgiannios13}. We note that for $\alpha \rightarrow \infty$, that is, when all energy is concentrated in a single-velocity component of $\sim 0.1$c, then the flux increases as $\propto t^3$ as the blast wave coasts. For $\alpha=0$, which means constant blast wave energy (no energy injection), one obtains the ``usual" temporal flux decay for a decelerating blast wave in the corresponding phases (e.g. \citealp{sarietal98, leventis2012, sironiandgiannios13}).  When the slowest component, with $\sim 0.1$c velocity, catches up with the mildly relativistic component, energy injection ceases and the flux decreases as $\propto t^{-3(1+p)/10}$, see equation (\ref{after}).  This occurs at $t_{\rm dec,SN}$ given by equation (\ref{tdec}). Therefore, due to our choice of the kinetic energy normalization of this mildly relativistic component, the radio SNR light curve will exhibit a modified light curve before $t_{\rm dec,SN}$, but the normalization at $t_{\rm dec,SN}$ and the light curve beyond this time will remain the same as presented in Section \ref{SNR_section}. 

As discussed above, the flux of the mildly relativistic component before $t_{\rm dec,SN}$ is given by $F_{\nu} = F_p (t/t_{\rm dec,SN})^{-s}$.  When the blast wave is non-relativistic and energy injection proceeds, then $s$ takes the value in equation (\ref{nonrelativistic}) and the flux (in $\mu$Jy) is given by
\beq
\begin{split}
F_{\nu} \approx  & \, 440 \, \bar{\epsilon}_{e,-1} \, \epsilon_{\text{B-SN},-2}^{\frac{1+p}{4}} \, \beta_{\rm SN,-2}^{\frac{40+11\alpha+p(\alpha-20)}{10+2\alpha}} \, E_{\rm SN,52.5}^{\frac{3+5p}{10+2\alpha}}  \, n_0^{\frac{p(\alpha-5)+5\alpha+19}{20+4\alpha}}\\
& \times\nu_{{\rm GHz}}^{\frac{1-p}{2}} \, (1+z)^{\frac{p(10-\alpha)-5\alpha-16}{10+2\alpha}} \, d_{\rm 27}^{-2} \, \Big(\frac{t}{29 \, {\rm yr}}\Big)^{-s},
 \end{split}
\label{F_NRMR}
\eeq
where we have left the redshift dependence since it is non-trivial. As an example, if we let $\alpha=5$ and $p=2.4$ this flux is given by (in $\mu$Jy)
\beq
F_{\nu}\approx 35 \,  \,  \bar{\epsilon}_{e,-1} \, \epsilon_{\text{B-SN},-2}^{0.85} \, \beta_{\rm SN,-2}^{2.95} \, E_{\rm SN,52.5}^{0.75} \, n_0^{1.1} \nu_{{\rm GHz}}^{-0.7} \, d_{\rm 27}^{-2} \, t_{\rm yr}^{0.75},
\eeq
which rises slower than $\propto t^3$.

It is also instructive to determine the time when the decaying flux of the GRB jet component (at this late time it is spherical and in the deep Newtonian regime) is equal to the flux of the SNR (before $t_{\rm dec,SN}$) as was done in \cite{barniolduranandgiannios15}. Setting equation (\ref{F_NRMR}) equal to the flux of the GRB afterglow (equation (7) in \citealp{barniolduranandgiannios15}), we obtain the time after which the SNR outshines the GRB afterglow (we will call this the rebrightening time, $t_{\rm rb}$). As an example, for $\alpha=5$ and $p=2.4$, the rebrightening time is
\beq
t_{rb} \approx (1.8 {\rm yr}) \, \beta_{\rm SN,-1}^{-\frac{5}{3}} \, E_{\rm SN,52.5}^{-0.42} \, E_{\rm GRB,51}^{0.76} \, n_{0}^{-\frac{1}{3}} \, \chi_B^{-0.48}.
\label{t_rb}
\eeq
In this expression, $E_{\rm GRB}$ is the true (beaming-corrected) energy of the GRB jet, and 
$\chi_B \equiv \epsilon_{\text{B-SN}}/\epsilon_B$. 

Taking $\alpha\rightarrow\infty$ (i.e. assume no mildly relativistic component) and $p = 2.4$ gives $t_{rb} \sim 9$ yr, which is significantly longer. Therefore, the presence of a mildly relativistic component in the SN ejecta naturally yields a radio SNR flux that exceeds the GRB component at a much earlier time. We note that to obtain equation (\ref{t_rb}), we have used the simplifying assumption that the GRB emission is in the DN phase and that the energy injection to the SN ejecta is in the non-relativistic phase. As for the SNR emission, we assume that this mildly relativistic SN-component is quasi-isotropic; therefore, it is observable for any viewing angle. 

The results presented in this subsection are valid for max($\nu_a,\nu_m$) $< \nu < \nu_c$. At early times, in the relativistic phase, the characteristic frequencies are larger; therefore the observing frequency could be either below $\nu_m$ or $\nu_a$. We use the expressions in Barniol Duran et al. (2015; see their equations 18 and 19) to estimate the location of the characteristic frequencies and the expected light curves. Whenever the characteristic frequencies cross the observing band, a sharp break in the radio light curves occurs when max($\nu_a,\nu_m$) $< \nu < \nu_c$. At later times, the light curve transitions to the non-relativistic phase, then to the DN phase, and then finally to the time when energy injection ceases (see Fig. \ref{fig2}).

\section{Summary of results} \label{Section_results}

In the following subsections we present the expected optical and radio light curves, respectively, including all different components discussed above.

\subsection{Optical emission} \label {Optical emission}

\begin{figure}
\includegraphics[width=9cm]{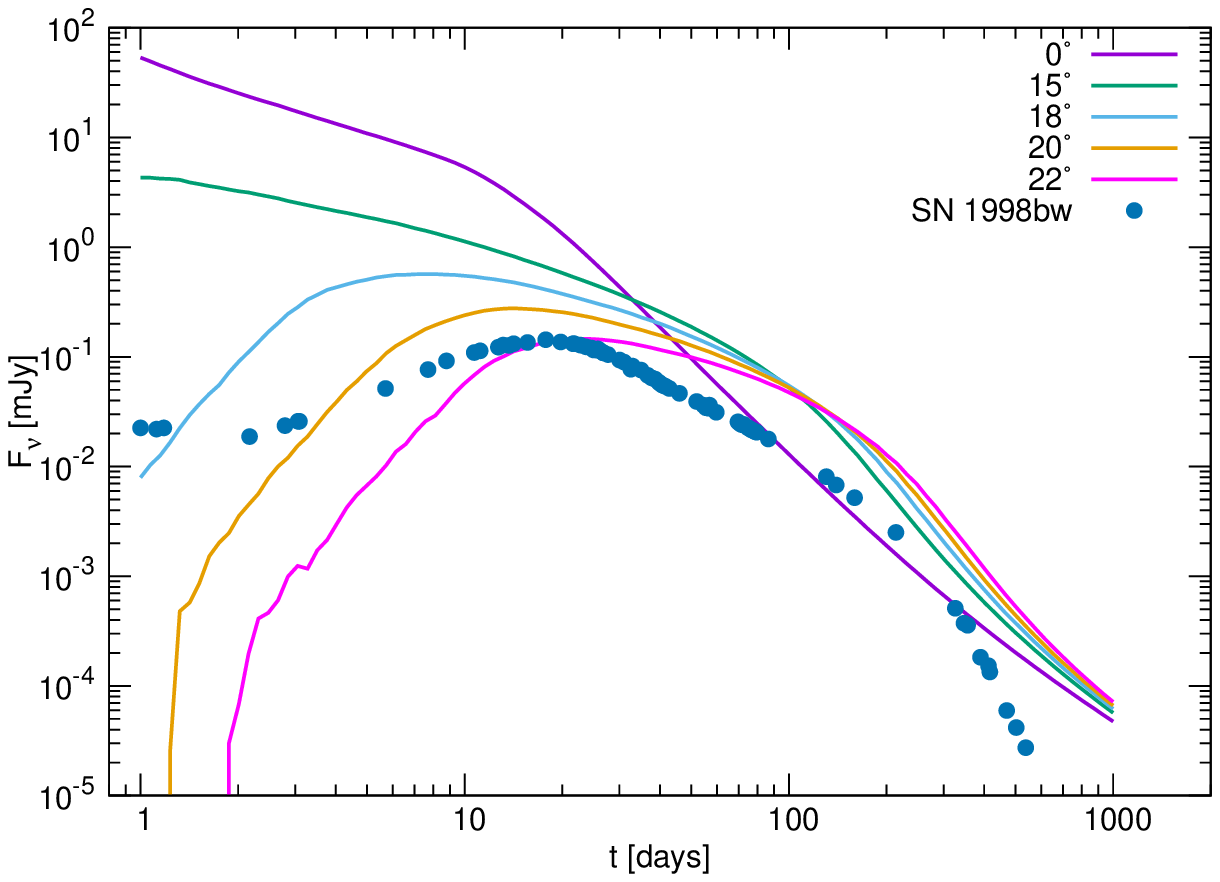} 
\includegraphics[width=9cm]{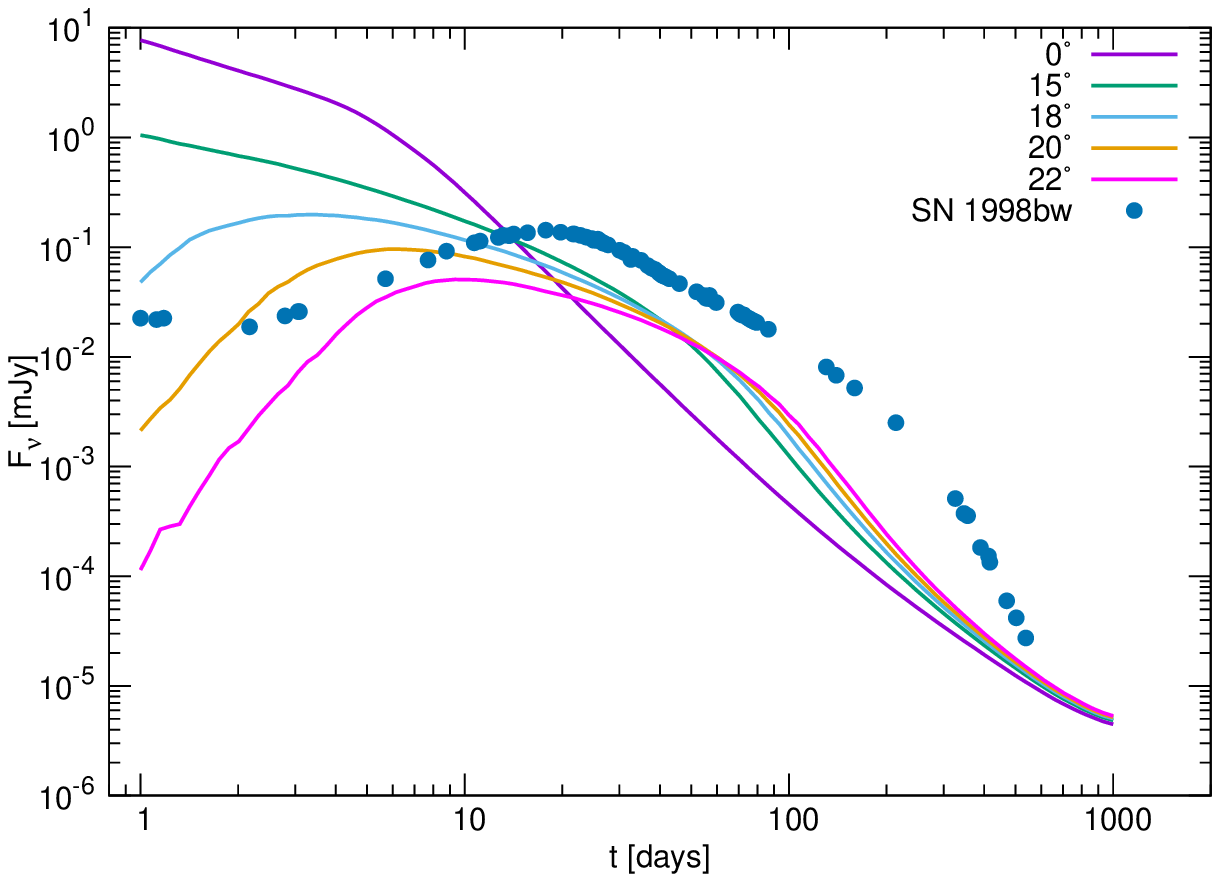}
\caption{Optical ($\sim 2$ eV) GRB afterglow light curves (lines) and the SN optical emission (points); for the latter we use observations of SN 1998bw, which serves as a typical GRB-accompanying SN. The top panel shows the model for the ``brightest" afterglows in our sample, while the bottom panel shows the model for the afterglow observed in the 5$^{th}$ decile. In both panels we show the on-axis and off-axis afterglows (viewing angles are indicated in the legend, peak flux of light curves decreases for larger viewing angles). The SN optical emission outshines the off-axis afterglow emission unless the viewing angle is very close to twice the half-opening angle of the GRB jet. An external density of $n = 1$ cm$^{-3}$ was used (for other parameters, see Section \ref{Baseline_optical}). The source is placed at a distance $d_{L}=300$ Mpc. Afterglow light curves are produced with the Afterglow Library (\protect\citealp{vaneertenandmacfadyen12}).}
\label{fig1}
\end{figure}

We have calculated the GRB optical afterglow for each decile for different viewing angles, $\theta_{\rm v}$.  We compare this emission with that of the optical emission of SN 1998bw.  As an example, the light curves for the ``brightest afterglow" and the 5$^{th}$ decile are shown in Fig. \ref{fig1} (we plot these for $\theta_j=0.2$). As can be seen, the SN optical emission is brighter than any off-axis afterglow for which the viewing angle is larger than $\theta_{\rm v,crit} \sim 20^{\circ} \sim 1.7 \theta_j$. A similar conclusion holds for $\theta_j=0.1$ (we obtain $\theta_{\rm v,crit} \sim 15^{\circ}$).

\subsection{Radio emission}

\begin{figure}
\includegraphics[width=9cm]{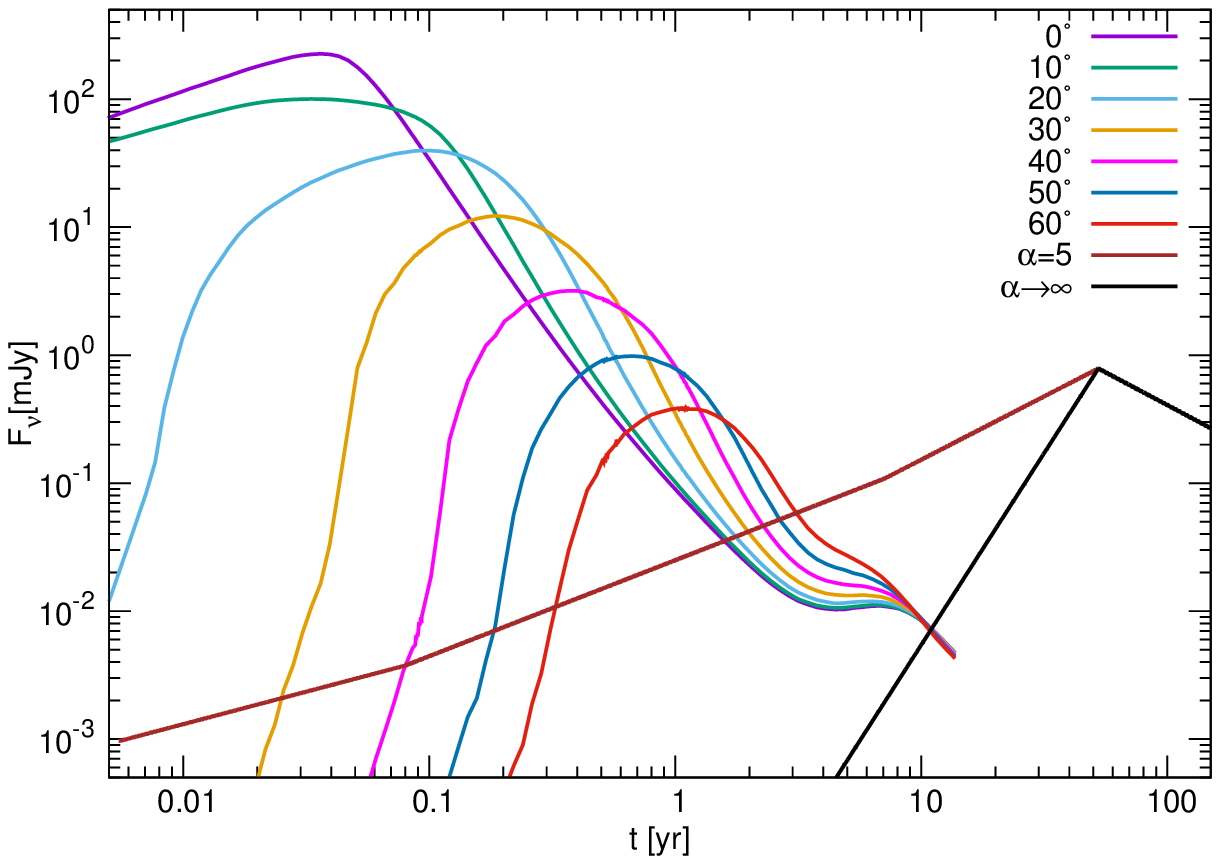} 
\includegraphics[width=9cm]{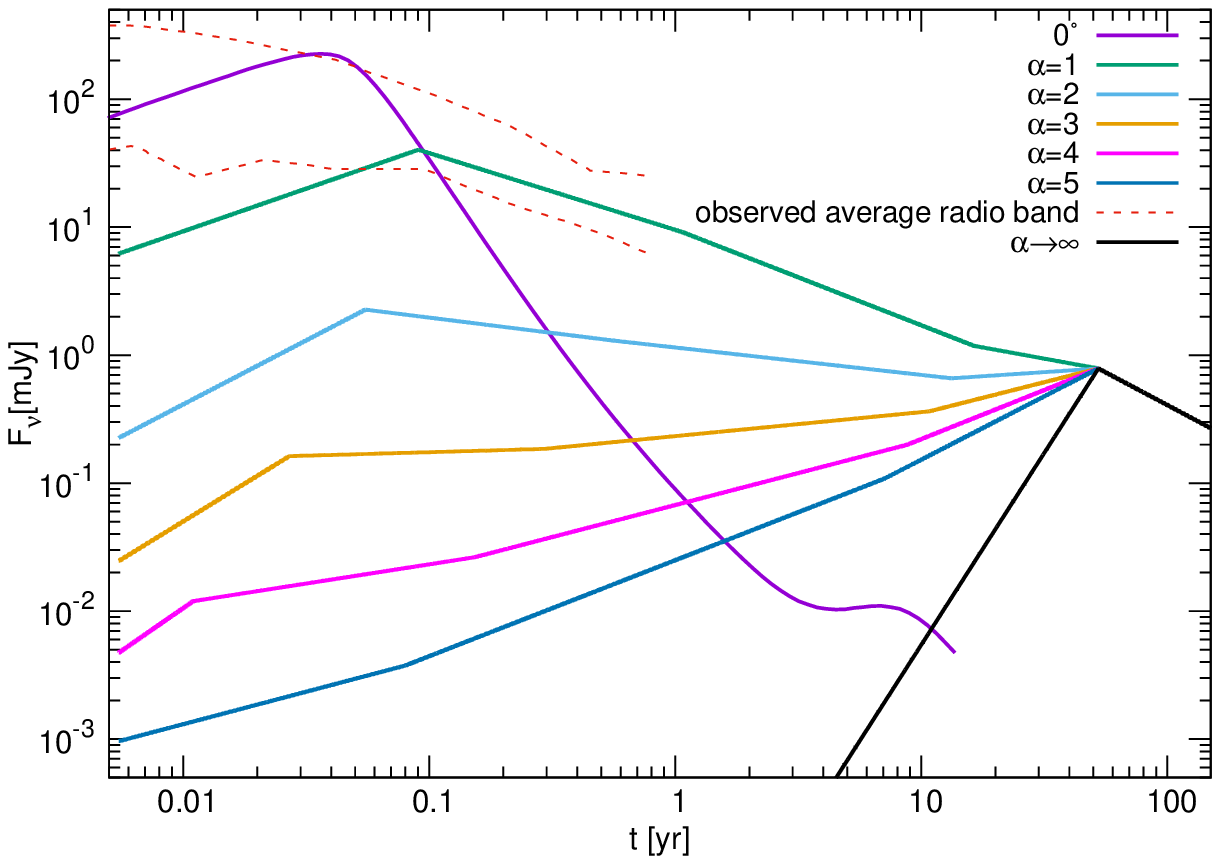}
\caption{Top panel: Radio (4.9 GHz) GRB afterglow light curves for the ``average" afterglow of our sample for various viewing angles (see legend, peak flux of light curves decreases for larger viewing angles). Radio SNR emission, allowing for the presence of a mildly relativistic component in the SN ejecta for different values of $\alpha$ (see legends in both panels, higher values of $\alpha$ correspond to lower fluxes at 1 yr), where the kinetic energy is injected to the blast wave as a function of velocity $\propto (\beta \gamma)^{-\alpha}$. Bottom panel: The region between the dashed lines indicates the location of the observed on-axis radio afterglows from the sample of \protect\cite{chandra2012}.
The mildly relativistic component generally shows four breaks in its light curve: the crossing of the minimum synchrotron frequency, the transition to the non-relativistic phase, the transition to the deep Newtonian phase, and the cessation of energy injection (in order of appearance, see Section \ref{Mildly_relativistic_section}). An external density of $n = 1$ cm$^{-3}$ was used (for other parameters, see Section \ref{Baseline_optical}), the source is placed at $d_{L}= 300$ Mpc. Afterglow light curves are produced with the Afterglow Library (\protect\citealp{vaneertenandmacfadyen12}). The SNR radio emission is calculated as in \protect\cite{barniolduranandgiannios15}.}
\label{fig2}
\end{figure}

Using the parameters obtained from matching the optical data at 1 day, we have calculated the GRB radio afterglow for each decile for different viewing angles.  As an example, the light curves for the ``average afterglow" are shown in the top panel of Fig. \ref{fig2}. We plot these for $\theta_j=0.2$. For $\theta_j=0.1$ the on-axis radio light curve peaks only a factor of $\sim 2$ earlier and with similar flux. The $\theta_j=0.1$ off-axis light curves peak at similar times, although with weaker fluxes by a factor of $\sim 4$, compared with the $\theta_j=0.2$ case.

We present the SNR radio emission, and also the possible contribution of a mildly relativistic SN component with $\alpha \approx 5$  (e.g., \citealp{tan2001}), in the top panel of Figure \ref{fig2}.  In the bottom panel of Figure \ref{fig2}, we only include the on-axis radio afterglow for the average afterglow, but include several possible distributions for the mildly relativistic SN component: $1 \le \alpha \le 5$, and $\alpha\rightarrow\infty$, which corresponds to the case where the SN ejecta has a single speed of $\sim 0.1$c (no mildly relativistic component). As can be seen, allowing for the presence of a mildly relativistic component yields a contribution to the radio flux at earlier times. 

Our predicted on-axis average radio light curve (predicted using the parameters obtained from matching the optical data at 1 day) does fairly well in reproducing the early ($<$ month) average radio observations in \cite{chandra2012}\footnote{As pointed out by Chandra \& Frail (2012), the bright radio data at $<$ few days might be the result of an extra component: the reverse shock emission, which should decay quickly afterward and will not affect our radio light curves at late times.}. However, it underpredicts the late time ($>$ months) radio observations. These late time data ($>$ months) in \cite{chandra2012} are sparse: their average observed radio afterglow light curve at this stage is dominated only by a few bright long-lived afterglows. However, it does seem that some specific radio afterglows do decay much slower than expected in the simplest external shock model (see, e.g., \citealp{panaitescu2004}). We discuss this in the next section.

\section{Discussion} \label{Discussion}

In Fig. \ref{fig1} we see that the brightest afterglows are visible up to a critical viewing angle of $\theta_{\rm v,crit} \sim 20^{\circ}$, beyond which the SN emission becomes comparable to the optical afterglow. For fainter optical afterglows, the SN emission outshines the afterglow at early times ($\la 1$ month), but it will still be detectable up to $\theta_{\rm v,crit}$.  This pattern is seen for afterglows up to the $5^{th}$ decile. For optical afterglows in the $\le 6^{th}$ decile, the critical viewing angle is $< 20^{\circ}$. Therefore, optimistically, $\sim 50$\% of afterglows in our sample outshine the optical SN emission as long as the viewing angle is $ < \theta_{\rm v,crit}$. From this, we can calculate the solid angle subtended by observers with viewing angle within $\theta_{\rm v,crit}$ and divide this by the total solid angle to find the probability of detecting such afterglows (including the counter jet). We find a small probability of $100\times 0.5 \int_ {0}^{\theta_{\rm v,crit}} sin{\theta} d{\theta} \approx 3$\% to clearly identify the afterglow emission in a GRB associated with a SN. If we consider $\theta_j=0.1$, this probability decreases to $\sim 2$\%. As mentioned in Section \ref{Baseline_optical}, there are different sets of parameters that could be used to match the optical on-axis fluxes. In the Appendix, we investigate the afterglow light curves in Fig. 1 for different sets of $E_{\rm iso}, n$ and $\epsilon_B$. We find that our conclusions do not change even when considering this degeneracy in the parameters.

In the radio band the afterglow is observable for various viewing angles and for long times as shown in Fig. \ref{fig2}. At late times, the radio SNR emission outshines the GRB afterglow \citep{barniolduranandgiannios15}. The rebrightening time decreases significantly if the SN ejecta contains a mildly relativistic component. Therefore, the time when the rebrightening occurs can give us important information on the energy distribution of the SN ejecta. The radio SNR emission does depend on the microphysical parameters of the SN shock. Here, we used microphysical parameters similar to those obtained in young radio SNe (e.g., \citealp{chevalier2006}, see discussion in \citealp{barniolduranandgiannios15}).

For a smaller density by a factor of 10 ($n=0.1$ cm$^{-3}$, increasing $\epsilon_B$ correspondingly, so that the optical flux at 1 day is matched), the on-axis radio afterglow peaks earlier by a factor of $\sim 1.5$ and has a similar flux as the $n=1$ cm$^{-3}$ case. The off-axis radio afterglows peak a factor of $\lae 4$ later in time with a factor of $\lae 3$ smaller fluxes.  Also, for this smaller density, the peak time of the SNR radio emission increases by a factor of $\sim 2$, which is expected from equation (\ref{tdec}). This causes the rebrightening time to increase by a factor of $\sim 2$. 

As can be seen in the bottom panel of Fig. \ref{fig2}, our on-axis GRB radio afterglow underestimates the late time ($>$ months) average observed radio flux.  Although the data used in \cite{chandra2012} to construct the late time average light curve are sparse, there {\it are} some bursts that show a shallow radio light curve. Its origin has been studied by \cite{panaitescu2004}, who have considered several possibilities (including time-varying microphysical parameters, and energy injection to the reverse shock, among others), but it is clear that it cannot have an external shock origin in the simplest model. The theoretically calculated radio on-axis afterglows decrease too fast.

\begin{figure}
\includegraphics[width=9cm]{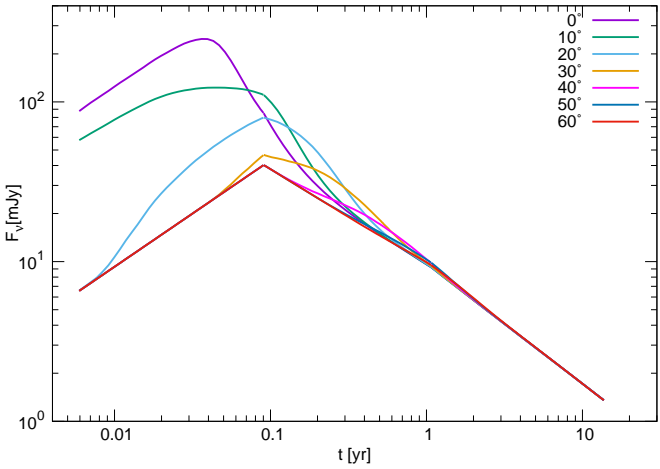} 
\caption{The sum of the ``average" radio (4.9 GHz) GRB afterglow light curves (for different viewing angles, see legend, peak flux of light curves decreases for larger viewing angles) and the radio emission from an $\alpha=1$ mildly relativistic component, which is quasi-spherical (see Fig. \ref{fig2}, bottom panel). An external density of $n = 1$ cm$^{-3}$ was used (for other parameters, see Section \ref{Baseline_optical}), the source is placed at $d_{L}= 300$ Mpc. Afterglow light curves are produced with the Afterglow Library (\protect\citealp{vaneertenandmacfadyen12}).}
\label{fig3}
\end{figure}

In order to account for the observed flat late-time radio curves, we consider a SN ejecta with a mildly relativistic contribution with $\alpha = 1$, which nicely follows the late-time on-axis radio afterglow of \cite{chandra2012}. Therefore, we can add both contributions to the radio band: 1. the on-axis GRB afterglow and 2. the (quasi-spherical) mildly relativistic SN ejecta component, to obtain the late time radio on-axis light curve.  The precise physical origin of the second component is outside of the scope of this paper (it could be, e.g., a mildly relativistic component from the GRB jet itself, and probably not related to the SN). Nevertheless, at these late times, the component is most likely non-relativistic and quasi-spherical, allowing observers at any angle to detect it. We present the contribution from both components for different viewing angles in Figure 3. In light of this, radio fluxes presented in the top panel of Fig. \ref{fig2} serve as lower limits.

After finding that the $\alpha=1$ component reproduces the observed late time on-axis afterglow radio data quite well, we investigated the contribution of this component in the optical band, i.e., its contribution to Fig. \ref{fig1}. We found that this component outshines the SN optical emission after $\sim 3$ months. However, for typical cosmological GRBs, which occur at $z \gae 1$, the optical flux from the $\alpha=1$ component is too weak to be observed. Nearby GRBs ($\lae$ 300 Mpc) might show a flattening in the optical light curve at $\sim 3$ months due to the contribution of this $\alpha=1$ component. 

As mentioned above, the vast majority of off-axis optical afterglows are expected to be weaker than the emission from the accompanying SN.  This can make the detection of off-axis optical afterglows a more difficult task than previously thought (e.g., \citealp{totani2002, rossi2008, ghirlanda2015}), even if they are stronger compared to their host galaxy emission. Optical off-axis afterglow studies should include the contribution from the SN emission, since the SN-GRB association is firm. 

Several radio surveys are coming online in the near future. We briefly mention their potential in detecting radio off-axis afterglows for ``average" GRBs in view of our results (these ``average" GRBs have isotropic equivalent energy $E_{iso}\sim10^{53}$ ergs).  We consider one of those programmes: VAST, which is the ASKAP Survey for Variables and Slow Transients (\citealp{murphy2013}). They have planned several surveys to detect radio transients at $\sim 1.4$ GHz. The ``VAST-Wide" survey has a 1$\sigma$ rms sensitivity of 0.5 mJy and covers $10^4$ deg$^2$ per day. From Fig. 2, we see that the peak flux at $\theta_{\rm v}\approx45^{\circ}$ is $\sim 1$ mJy for a frequency of 4.9 GHz (the flux at 1.4 GHz would be a factor of $\sim 2$ larger). Therefore we deduce that this survey will be able to detect afterglows up to a distance of 300 Mpc provided $\theta_{\rm v}\lae45^{\circ}$.  With the covered area of the survey and using the beaming-corrected local GRB rate of\footnote{The local GRB rate is $\sim 1$ Gpc$^{-3}$ yr$^{-1}$ (\citealp{wanderman2010}) for events with gamma-ray isotropic luminosities $\gae 10^{50}$ erg/s. For the bursts energies considered in Fig. \ref{fig2}, the local rate would be a factor of $\sim 3$ smaller (see \citealp{metzger2015}).  For $\theta_j = 0.2$, the beaming correction is $ 2 \pi \theta_j^2 / (4 \pi) \sim \frac{1}{50}$ (e.g., \citealp{guetta2005}). Therefore, the beaming-corrected local rate is $\sim 15$ Gpc$^{-3}$ yr$^{-1}$.} $\sim 15$ Gpc$^{-3}$ yr$^{-1}$, we expect only $\sim 0.4$ events per year. The ``VAST-Deep Multi-field" survey has a 1$\sigma$ rms sensitivity of 0.05 mJy and covers $10^4$ deg$^2$ per year. Constraining $\theta_{\rm v}\lae45^{\circ}$, we expect to detect $\sim 13$ events per year.  Finally, the ``VAST-Deep Single field" survey has a 1$\sigma$ rms sensitivity of 0.05 mJy but covers $30$ deg$^2$ per day, which yields $\sim 0.04$ events per year (here again we keep the same constraint on $\theta_{\rm v}$). Since these surveys will last for a $\sim$ few years, off-axis radio afterglows should be detected (see \citealp{ghirlanda2014, metzger2015}). Other radio surveys such as Apertif on WSRT (\citealp{oosterloo2010}), MeerKat (\citealp{booth2009}), surveys with the VLA (\citealp{perley11}) known as VLASS, SKA1 (\citealp{carilli04}), should also be able to detect orphan afterglows, provided that it is possible to distinguish between them and other slowly evolving radio synchrotron sources.

The calculations in the previous paragraph take into account the afterglow light curves of Fig. 2, which were shown to underestimate the radio flux after a few months. If we consider the presence of a mildly-relativistic quasi-spherical component (see Fig.~3), the radio flux of the source for any observing angle is increased by one order of magnitude. Thus even if a modest fraction of GRBs contain such a mildly relativistic component, the number of orphan afterglows detected in the radio would have a significant increase. 

Since the nearby population of GRBs is likely to be dominated by low-luminosity GRBs ({\it ll}GRBs, e.g., \citealp{soderberg2004}b), we briefly discuss the detection of {\it ll}GRBs in regards to upcoming surveys. Even though they are more abundant and roughly isotropic, the afterglows of {\it ll}GRBs are a couple of orders of magnitude fainter than off-axis long GRBs. This makes them quite hard to detect in upcoming radio surveys; SKA1 is the only survey that will be able to detect them (\citealp{metzger2015}). Nevertheless, detection and very late time follow-up of the radio afterglows of {\it ll}GRBs should give us insight into the energy distribution of their blast waves and the emergence of the radio SNR (see \citealp{barniolduran15,barniolduranandgiannios15}). 

\section{Conclusions} \label{Conclusions}

We have explored the different components present in a long GRB explosion: the GRB jet, the SN ejecta and the possible contribution of a mildly relativistic component.  The GRB jet interacts with the external medium and produces an afterglow, which can in principle be detected for different viewing angles without the detection of the prompt gamma-ray emission.  The SN ejecta is generally quasi-spherical and produces optical photons detected by observers at any angle.  

Using a sample of optical on-axis afterglows, we predict that the vast majority of optical off-axis afterglow will be too weak to be detected in excess of the emission from the accompanying SN. In lines with previous work, radio observations provide the best alternative to detect afterglow without an associated gamma-ray trigger.  

Upcoming radio surveys should be able to detect off-axis afterglows within $\sim 300-500$ Mpc.  These radio surveys can act as radio ``triggers". We encourage the follow-up of newly found off-axis afterglows with dedicated facilities (e.g., VLA). Follow-ups may discover the emission from the radio SNR that accompany long GRBs (\citealp{barniolduranandgiannios15}). In addition, they can strongly constrain the energetics of any mildly relativistic component (either associated with the SN ejecta or the GRB jet).  Finally, the discovery of nearby GRB sources could help us in identifying their central engines. For example, bright soft gamma-ray repeater-like flares at a location coincident with that of the afterglow could reveal that the GRB is associated with the birth of a magnetar (\citealp{giannios2010}).

\section*{Acknowledgements}

We thank Laura Chomiuk for useful discussions, and Alexander van der Horst and Maria Petropoulou for providing useful comments on the draft. We acknowledge support from NASA through grant NNX16AB32G issued through the Astrophysics Theory Program. We also acknowledge support from the Research Corporation for Science Advancement's Scialog program. Development of the Boxfit code was supported in part by NASA through grant NNX10AF62G issued through the Astrophysics Theory Program and by the NSF through grant AST-1009863.

\bibliographystyle{mn2e}
\bibliography{references} 

\section*{Appendix}

\begin{figure}
\includegraphics[width=9cm]{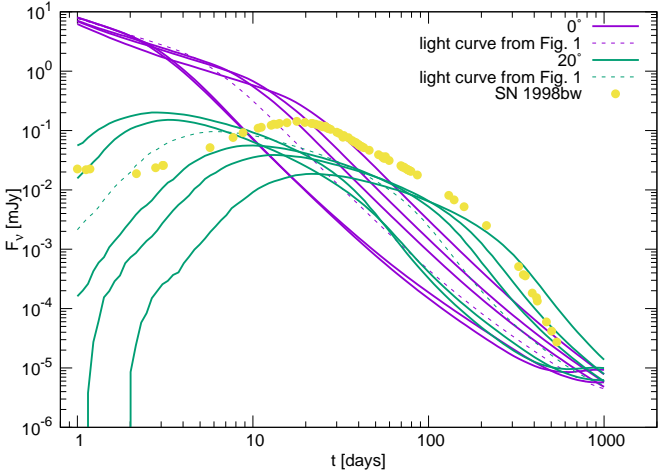} 
\caption{Optical ($\sim 2$ eV) GRB afterglow light curves (lines) both on-axis ($0^{\circ}$, these have a flux of $\sim 5$ mJy at 1 day) and off-axis ($20^{\circ}$) for multiple parameters (see Table \ref{table1}) and the SN optical emission (points); for the latter we use observations of SN 1998bw. The GRB afterglow light curves are modelled from afterglow observations in the 5$^{th}$ decile. Dashed lines correspond to the light curves in the bottom panel of Fig. 1 (its parameters are in italics in Table 1). The SN optical emission outshines the off-axis afterglow emission unless the viewing angle is very close to  twice the half-opening angle of the GRB jet ($\theta_{j}=0.2$). We fix $\epsilon_{e}=0.1$, source is placed at a distance $d_{L}=300$ Mpc. Afterglow light curves are produced with the Afterglow Library (\protect\citealp{vaneertenandmacfadyen12})}
\label{fig4}
\end{figure}

In Section \ref{Baseline_optical} we considered a set of parameters that matches the optical flux at 1 day for each decile. These parameters were used to calculate the off-axis optical light curves and to estimate an approximate critical angle, $\theta_{\rm v,crit}$ (see section \ref{Optical emission}). Here we show that our conclusions on $\theta_{\rm v,crit}$ remain approximately the same when we consider different sets of parameters that also match the optical flux at 1 day. We allow $n$, $E_{iso}$ and $\epsilon_{B}$ to vary in a large range (see Figure \ref{fig4} and Table \ref{table1}) while fixing $\epsilon_{e}=0.1$ and $\theta_{j}=0.2$. For the parameters considered in this Appendix, $\theta_{\rm v,crit}$ remains approximately the same until the 5$^{th}$ decile. For lower deciles, $\theta_{\rm v,crit}$ is smaller since the on-axis optical afterglow emission is weaker, just as discussed in Section \ref{Discussion}.

\begin{table}
\begin{center}
\begin{tabular}{cccc}
\hline
$\epsilon_B$ & n [cm$^{-3}$] & $E_{\rm iso}$ [erg] & $\eta$ [\%]\\
\hline
$2\times10^{-4}$ & $0.1$ & $9\times10^{53}$ & $4$\\
$3\times10^{-3}$ & $0.1$ & $2\times10^{53}$ & $20$\\
$9\times10^{-5}$ & $1$ & $7\times10^{53}$ & $5$\\
$\mathit{8\times10^{-4}}$ & $\mathit{1}$ & $\mathit{2\times10^{53}}$ & $\mathit{20}$\\
$1\times10^{-2}$ & $1$ & $4\times10^{52}$ & $100$\\
$2\times10^{-4}$ & $10$ & $2\times10^{53}$ & $20$\\

\hline
\end{tabular}
\end{center}
\caption{Different parameters considered in Figure \ref{fig4} that match the on-axis optical flux at 1 d for the 5$^{th}$ decile. The rows are arranged in the order of descending flux of the light curves (in Figure \ref{fig4}) at 200 days. For example, the first row corresponds to the  parameters of the light curve with the brightest flux at 200 days (at $0^{\circ}$ and $20^{\circ}$). The row in italics indicates the set of parameters used for the afterglow light curves in the bottom panel of Figure \ref{fig1} (dashed lines in Figure \ref{fig4}). } 
\label{table1}
\end{table}

\end{document}